# Unveiling the Interfacial Reconstruction Mechanism Enabling Stable Growth of the Delafossite $PdCoO_2$ on $Al_2O_3$ and $LaAlO_3$

Anna Scheid[1], Tobias Heil[1], Y. Eren Suyolcu[1], Qi Song[2], Niklas Enderlein[3], Arnaud P. Nono Tchiomo[4], Prosper Ngabonziza[4,5], Philipp Hansmann[3], Darrell G. Schlom[2,6,7], and Peter A. van Aken[1]

[1]Max Planck Institute for Solid State Research, Stuttgart, 70569, Germany

[2]Department of Materials Sciences and Engineering, Cornell University, Ithaca, New York 14853, USA

[3]Department of Physics, Friedrich-Alexander-Universität Erlangen-Nürnberg (FAU), 91058, Erlangen Germany

[4]Department of Physics and Astronomy, Louisiana State University, Baton Rouge, Louisiana 70803, USA

[5]Department of Physics, University of Johannesburg, P.O. Box 524 Auckland Park 2006, Johannesburg, South Africa

[6]Kavli Institute at Cornell for Nanoscale Science, Ithaca, New York 14853, USA

[7]Leibniz-Institut für Kristallzüchtung, Berlin, 12489, Germany

# Abstract

Delafossites, comprised of noble metal ($A^+$) and strongly correlated sublayers ($BO_2^-$), form natural superlattices with highly anisotropic properties. These properties hold significant promise for various applications, but their exploitation hinges on the successful growth of high-quality thin films on suitable substrates. Unfortunately, the unique lattice geometry of delafossites presents a significant challenge to thin-film fabrication. Different delafossites grow differently, even when deposited on the same substrate, ranging from successful epitaxy to complete growth suppression. These variations often lack a clear correlation to obvious causes like lattice mismatch. Unidentified stabilization mechanisms appear to enable growth in certain cases, allowing these materials to form stable thin films or act as buffer layers for subsequent delafossite growth.

This study employs advanced scanning transmission electron microscopy techniques to investigate the nucleation mechanism underlying the stable growth of $PdCoO_2$ films on $Al_2O_3$ and $LaAlO_3$ substrates, grown via molecular-beam epitaxy. Our findings reveal the presence of a secondary phase within the substrate surface that stabilizes the films. This mechanism deviates from the conventional understanding of strain relief mechanisms at oxide heterostructure interfaces and differs significantly from those observed for Cu-based delafossites.

# Introduction

Delafossites, with the general chemical formula $ABO_2$, consist of metallic ions ($A^+$) and $B^{3+}$ ions. Their unique crystal structure features alternating layers of $A^+$ ions sandwiched between planes of edge-sharing $[BO_6]^-$ octahedra[1,2]. Delafossite oxides exhibit a wide range of physical properties, including *p*-type or ambipolar semiconductivity (for $A$ = Cu, $B$ = Al, Cr, Fe, Ga, Y, In), photocatalytic activity ($AgGaO_2$), multiferroicity ($CuFeO_2$), and metallic conductivity ($PdCoO_2$, $PdCrO_2$, $PdRhO_2$, $PtCoO_2$, and $AgNiO_2$)[3-7]. Among these, metallic delafossites stand out due to their exceptionally high in-plane conductivity. For instance, $PtCoO_2$ boasts the highest conductivity per carrier of any known oxide material, and $PdCoO_2$ exhibits the longest low-temperature mean free path among all metals[8-11]. The conductivity of $PdCoO_2$ surpasses that of pure Pd metal by a factor of four at room temperature, approaching the values observed in elements like Cu, Ag, and Au[9,12]. This remarkable conductivity probably arises from the layered structure, where insulating, strongly correlated transition-metal oxide layers alternate with conductive, triangular coordinated Pd or Pt layers[13]. This arrangement creates intrinsic heterostructures with highly anisotropic properties and quasi-2D electrical conduction. The resistivity within the *ab*-plane of metallic delafossites is typically orders of magnitude lower than along the *c*-axis, as the conductivity primarily originates from the Pd/Pt $4d/5d$ orbitals[8,12]. The highly dispersive band associated with these states crosses the Fermi level, resulting in a cylindrical Fermi surface that reflects the quasi-2D nature of the electrons within the Pd/Pt layers[12]. In addition, delafossites are chemically stable up to temperatures of 800-925 °C, allowing for applications in the high-temperature regime[1].

Given the limited size of delafossite single crystals (typically a few millimeters) and the growing interest in their applications in electronics, spintronics, and oxide heterostructures, significant effort has been devoted to the production of metallic delafossite thin films using techniques such as reactive sputtering, pulsed-laser deposition (PLD), and molecular-beam epitaxy (MBE)[14,15]. Unfortunately, the electrical conductivity of these thin films has not yet reached the levels observed in single crystals. While room-temperature conductivity can be comparable, low-temperature resistivity often differs by orders of magnitude[11,16].

The choice of substrate is crucial for successful thin-film growth. In general, the lattice mismatch between the film and substrate can induce strain, degrade properties, and introduce defects that diminish film quality. Since delafossites possess trigonal crystal structures, they are commonly grown on the planes of substrates with 3-fold or 6-fold symmetry, such as $Al_2O_3$ (001) or (111) planes of cubic crystals[14,15]. Surprisingly, despite the often significant lattice mismatch between metallic delafossites and substrates (e.g., ~6% for $PdCrO_2$ on *c*-$Al_2O_3$), high-quality *c*-axis-oriented thin films can be successfully grown[16-19]. Moreover, some delafossites exhibit consistently good or poor growth behavior, regardless of the degree of lattice mismatch[20]. Intriguingly, some films with larger mismatches demonstrate better growth than those with smaller mismatches. These observations suggest the existence of an unknown mechanism that relieves lattice mismatch strain at the interface of certain delafossites, beyond the scope of conventional coherent and semi-coherent interface models. To gain a deeper understanding of these phenomena, it is crucial to investigate the interface structure between the substrate and the epitaxial films. Scanning transmission electron microscopy (STEM) techniques offer valuable insights into these interfacial interactions.

In this work, we utilize electron ptychography, electron energy loss spectroscopy (EELS), and conventional STEM imaging to elucidate the atomic-scale reconstruction at the interface of $PdCoO_2$ grown on $Al_2O_3$ (001) and $LaAlO_3$ $(111)_{pc}$, where the subscript denotes pseudocubic indices, via MBE. This investigation reveals an unconventional growth mechanism driven by a complex interplay between Al and Co at the interface. These findings have significant implications for improving the stable growth of delafossite thin films.

# Materials and methods

## Sample preparation

Thin films of $PdCoO_2$ were synthesized on (001)-oriented sapphire ($Al_2O_3$), $(111)_{pc}$-oriented $LaAlO_3$, and (111)-oriented $SrTiO_3$ substrates using reactive oxide MBE in a Veeco GEN10 MBE system. The substrates were heated to temperatures ranging from 500 °C to 580 °C, as measured by a thermocouple positioned near the substrate heater. During deposition, a gas mixture comprising approximately 80% ozone and 20% oxygen was introduced, with a background pressure ranging from $5 \times 10^{-6}$ to $8.5 \times 10^{-6}$ Torr. Shutter-controlled layer-by-layer growth of the $PdCoO_2$ thin films grown on $Al_2O_3$ substrates was achieved by actuating the MBE shutters to supply monolayer doses of Pd and Co, following the sequence of atomic layers along the $c$-axis of the $PdCoO_2$ crystal structure. $PdCoO_2$ films grown on $LaAlO_3$ and $SrTiO_3$ substrates were codeposited by simultaneously exposing the substrate to Co, Pd, and ozone molecular beams under conditions where the excess Pd supplied desorbs as PdO (g). After growth, the films were immediately cooled to 300 °C in the same ozone background pressure in which they were grown. Further details and growth parameter optimization are provided in a previous study[18].

Epitaxial films of $PdCoO_2$ were also deposited by pulsed laser deposition (PLD). These $PdCoO_2$ films were deposited on the (001)-oriented $Al_2O_3$ at a substrate temperature of 700 °C under an oxygen pressure of 0.1−0.15 Torr in the PLD chamber. To achieve a stoichiometric composition in the $PdCoO_2$ films, we employed the PLD growth method of alternately ablating the $PdCoO_2$ and mixed-phase $PdO_X$ targets[21].

## Thin film characterization

The thin film structure was characterized using a PANalytical Empyrean X-ray diffractometer with Cu K$\alpha_1$ radiation.

## STEM specimen preparation

Electron-transparent specimens were prepared by diamond cutting and tripod wedge polishing. Final thinning was achieved using a precision ion polishing system (Gatan Inc., PIPS II, Model 695) equipped with a liquid nitrogen cooling stage utilizing argon ions.

## STEM investigations

Scanning transmission electron microscopy (STEM) investigations were performed using a JEOL JEM-ARM200F equipped with a cold field emission gun and a probe Cs corrector (DCOR, CEOS GmbH). Measurements were conducted at ambient temperatures with an acceleration voltage of 200 kV. To enhance the signal-to-noise ratio, mitigate scan artifacts, and minimize sample drift effects, STEM images were generated from multi-frame acquisitions, utilizing high scanning speeds and applying post-acquisition cross-correlation.

For spectroscopic measurements, EELS data were acquired with a Gatan GIF Quantum ERS imaging filter, using a 5 mm entrance aperture and a camera length of 1.5 cm yielding a collection semi-angle of 111 mrad. Principal component analysis (PCA) was employed to enhance the signal-to-noise ratio, utilizing 15 principal components for accurate elemental mapping[22].

For comprehensive 4D STEM data acquisition, a MerlinEM Medipix3 detector by Quantum Detectors was utilized. The detector was operated in 1-bit mode, enabling optimal detection of one electron per pixel in a single frame with a rapid pixel dwell time of 48 µs, corresponding to a frame rate of 20833 fps. All 4D STEM data acquisitions were performed with a camera length of 80 cm and a probe convergence semi-angle of 20.4 mrad. Ptychographic reconstructions were carried out using ptychoSTEM, an open-source MATLAB script repository available on GitLab[23–25].

## Density functional theory calculations

Density functional theory (DFT) calculations were performed using the Quantum ESPRESSO suite[26–28]. Details regarding the calculations are provided in the Supplementary Material.

# Results and discussion

The PdCoO$_2$ thin films grown on Al$_2$O$_3$ and LaAlO$_3$ substrates demonstrate high structural quality, as confirmed by multiple complementary characterization techniques. X-ray diffraction (XRD) patterns reveal only the 00ℓ reflections corresponding to the bulk PdCoO$_2$ crystal structure, indicating that the films are phase-pure, epitaxial, and oriented with the *c*-axis perpendicular to the plane of the substrate. This conclusion is further substantiated by the presence of sharp reflection high-energy electron diffraction (RHEED) spots, signifying a well-ordered surface structure, and by low-magnification high-angle annular dark-field (HAADF) STEM images, which verify the uniformity of the film morphology (Supplementary Figure S1). Notably, the observation of Laue oscillations around the XRD reflections for PdCoO$_2$ films on Al$_2$O$_3$ provides direct evidence of atomically smooth surfaces, uniform film thicknesses, and well-defined film-substrate interfaces.

Figure 1 shows HAADF STEM images of the PdCoO$_2$ thin films along the [210] and [010] orientations on both substrates. Delafossites grow with a 30° rotation relative to the Al$_2$O$_3$ substrate orientation due to the lattice geometry; hence, we will consistently indicate the film orientations in STEM images. In HAADF images, contrast is generally proportional to the atomic number; thus, columns of Al and O atoms are either not at all or only very weakly visible, whereas the heavy La and Pd atomic columns appear the brightest, followed by the slightly lighter Co species. Along the [210] orientation, the oxygen octahedra lie in projection and discrete octahedral orientations cannot be distinguished.

The STEM cross-section images along the [010] orientation of the films reveal domains with different orientations of the CoO$_6$ octahedra in subsequent layers of the films on both substrates, indicated by ticks in the insets. The stacking faults and twins observed in the thin films occur because both configurations appear energetically equally favorable during growth[29]. These defects are likely the main reason for the lower residual resistivity ratio of thin films compared to bulk crystals[15].

In-plane rotational twins are present in all epitaxial delafossite thin films grown on non-delafossite *c*-axis oriented substrates, studied up to date, and are also visible in the XRD measurements of our films (Supplementary Figure S1)[17,29,30]. In addition to the twinning, which can be avoided using high-miscut substrates, we occasionally observe individual defects, such as steps on substrate terraces or point defects[20]. Nevertheless, we do not observe any impurity phases, such as PdO$_X$ or Co$_3$O$_4$, which often occur under improper growth conditions[16,31].

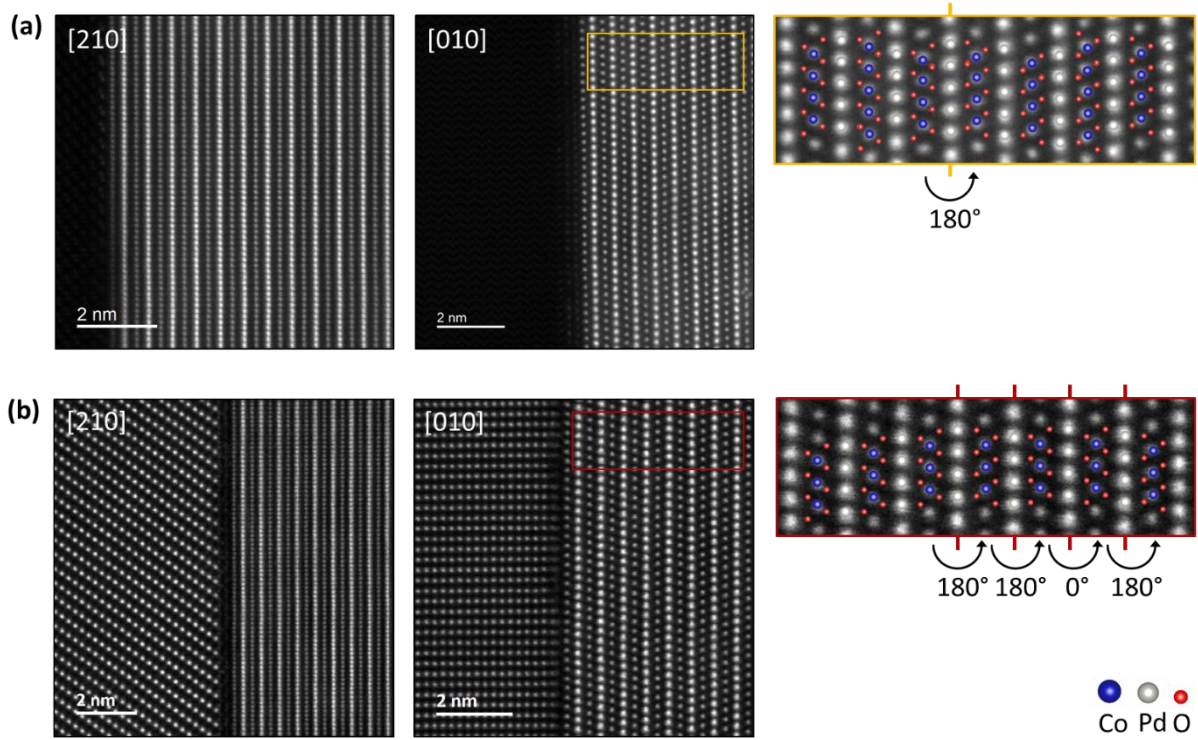

Figure 1. HAADF STEM images of PdCoO$_2$ thin films (right) on (a) Al$_2$O$_3$ and (b) LaAlO$_3$ for two structural orientations. For stable growth, the growth is always initiated with a [CoO$_2$]$^-$ layer. The insets in the [010] images reveal domains with opposing orientations of the CoO$_6$ octahedra in subsequent [CoO$_2$]$^-$-layers, and highlight stacking faults (0° swap) and twin domains (180° swap) in the films.

For a closer examination of the atomic structure at the interface, HAADF images, which consist of electrons mainly scattered to higher angles, do not provide sufficient contrast in the presence of light elements. Therefore, imaging techniques that are effective at visualizing oxygen are required. One of the most effective techniques for visualizing light elements is annular bright-field (ABF) imaging[32]. Here, an annular detector captures electrons only from the outer area of the bright-field disk where the contribution of beam interactions with light elements is relatively low due to channeling effects, resulting in low image intensities at the column locations.

Figure 2 shows simultaneously acquired STEM HAADF and ABF images of the thin film interface along two zone axes of a PdCoO$_2$ film grown on Al$_2$O$_3$ (Figure 2(a) and (b)) and a PdCoO$_2$ film grown on LaAlO$_3$ (Figure 2(c) and (d)). While the HAADF images provide strong contrast only for the heavier atomic species, in the simultaneously captured ABF images, the light elements are clearly visible. Figure 2(b) and (d) distinctly show the CoO$_6$ octahedra orientation, which is mirrored in the fourth (Figure 2(b)) and fifth (Figure 2(d)) layers due to their equal growth possibility.

The STEM images reveal that all the films are always initiated with a [CoO$_2$]$^-$ layer. MBE synthesis has shown that the films do not grow stably and tend to completely decompose if the growth is initiated with a Pd$^+$ layer[16]. This already indicates the importance of initiating film growth with cobalt in an ozone background pressure to stabilize the delafossites on the substrates. To test the likelihood of different interface types, we performed density functional theory (DFT) calculations using the Quantum ESPRESSO suite for the $PdCoO_2/Al_2O_3$ interfaces[26–28]. The stability of different interfaces was estimated by comparing the total energy of the (optimized) interface and that of its two constituents (film and substrate). We find that a Co-interface is significantly more stable (by 4 eV per unit cell area) compared to a Pd interface. More details are provided in the Supplementary Material.

Detailed examination of the insets in the ABF images reveals atomic columns at the interface that do not align with the structural model of the substrate and film along both orientations for the two different substrates. The unmatched atomic columns appear within the top layers of the substrates, with the films aligning perfectly to the initial $CoO_6$ octahedral layers. No structural defects, such as misfit dislocations, are observed within the first layer of the films. As is often observed in oxide heterostructures, chemical intermixing at the interface may occur, stabilizing the films on the substrate[33–35]. Alternatively, vacancies or interstitial sites within the substrate might be occupied during growth, contributing to the stabilization of the films[36,37]. Therefore, moving on from the Co-interface, we used DFT to examine the energetic feasibility of substituting Co-atoms into the first Al layer (next to the interface) of $Al_2O_3$. We calculated different cases of 50% Co-substitution (affecting two of the four Al-atoms per Al-layer) and the 100% case. The calculated binding energies exhibit a clear hierarchy: the 0% substitution (i.e., 100% Al) is the most stable, the 100% substitution (i.e., 100% Co) is the least stable, and the three tested cases of 50% Co-substitution settle between these extremes. While Co-substitution is not inherently energetically favorable, it is also not exceptionally unfavorable for the 50% substitution cases. This suggests that a smaller percentage of Co substitutions could statistically occur in experiment, particularly because the substrate will not achieve a perfectly Al-terminated surface despite pretreatment, and surface roughness likely results in numerous Al-vacancies. Furthermore, we constructed interfaces involving a Co-interface, where additional Co has been incorporated into the stoichiometrically induced Al-vacancies of $Al_2O_3$ as the unmatched atomic columns in the inset of Figure 2(a) seem to imply. Under the condition that the lattice spacing of the $Al_2O_3$ substrate was constrained during all geometry optimizations, the incorporation of additional Co within the $Al_2O_3$ led to a seemingly unphysical enhancement of the bonding distance between the Co-interface and the oxygen connected to the Co-inserted Al-layer of $Al_2O_3$. This unexpected bond elongation along the $c$-axis occurs as a compensatory response to the in-plane compressive strain induced by Co insertion, suggesting that the incorporation of Co into the $Al_2O_3$ lattice is not energetically favorable without structural reconstructions.

Therefore, the unmatched atomic columns observed in the ABF images suggest a whole atomic reconstruction at the interface between the substrates and the films. While an increased contrast in the HAADF images has been noted in other studies of $PdCoO_2$ thin films on $Al_2O_3$, the emergence of such an interfacial phase within the substrate surface has not been observed for $PdCoO_2$ thin films before and may be a crucial cornerstone for the stable epitaxial growth[17,20].

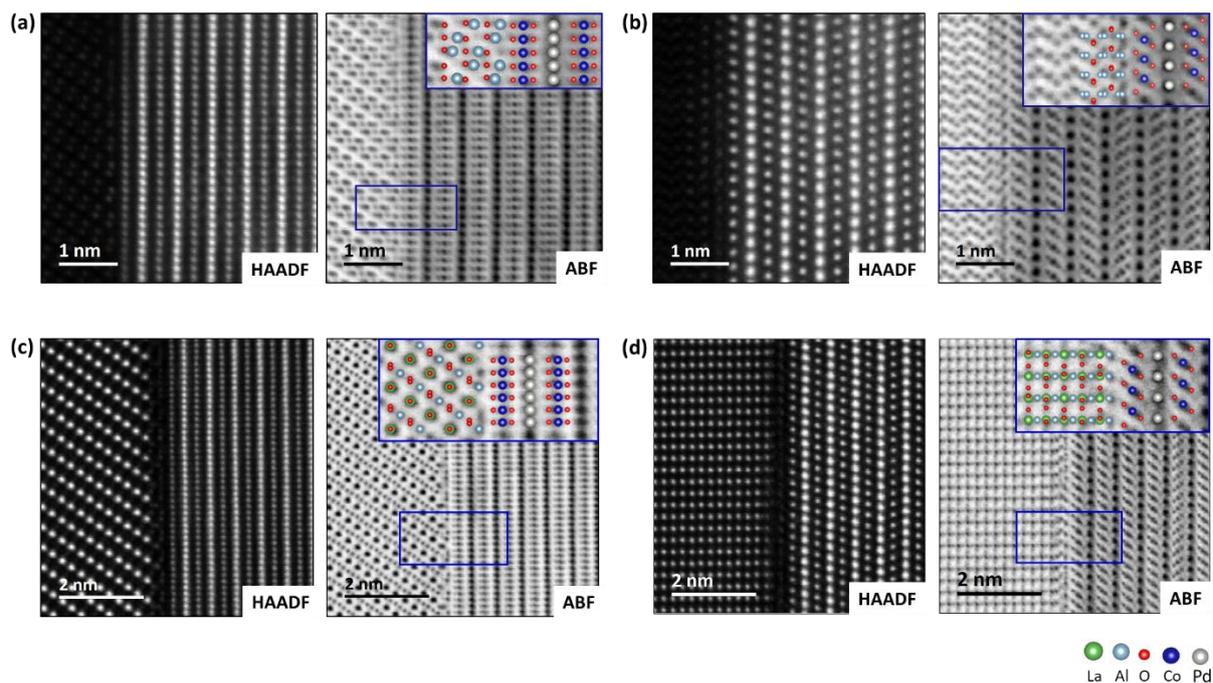

Figure 2. Simultaneously acquired STEM HAADF and ABF images of the $Al_2O_3/PdCoO_2$ interface for the (a) [210] and (b) [010] zone axes and the $LaAlO_3/PdCoO_2$ interface for the (c) [210] and (d) [010] zone axes. While the HAADF STEM images show strong contrast primarily for the heavy elements Pd, Co, and La, the ABF images clearly reveal the oxygen and aluminum columns. The insets in the ABF images show an overlay with the structural models of the substrates and thin films with unmatched atomic contrast at the interface within the substrates.

To characterize the chemical nature of the phase involving the unmatched atomic columns, we performed electron energy loss spectroscopy (EELS) across the interface. Previous studies have demonstrated that reliable interpretations of interface structures are attainable only under ultrathin specimen conditions ($\leq$ 20 nm)[38,39]. This minimizes misinterpretations arising from thickness effects and complex propagation phenomena such as beam broadening, crosstalk, and de-channeling. Therefore, we selected a thin sample area of the specimen for atomic-scale quantification of the interface structure. The thickness was estimated as 8.3 nm for the $PdCoO_2/Al_2O_3$ interface and 11.4 nm for the $PdCoO_2/LaAlO_3$ interface. This was achieved by comparing measured PACBEDs from a 4D STEM scan in the same sample area to multislice simulations of PACBED patterns using the *abtem* algorithm (Supplementary Figure S3)[40].

EELS data acquired from the different delafossite samples enabled the isolation of signals from characteristic energy-loss edges, allowing for a detailed analysis of the elemental distribution at the interface. Figure 3(a) and (c) present the extracted signals from the characteristic energy-loss edges of O (O K edge), Al (Al K edge), Co (Co $L_{3,2}$ edge), La (La $M_{5,4}$ edge), and Pd (Pd $M_{5,4}$ edge). The white arrows indicate the physical boundary between substrate and thin film. The STEM-EELS elemental maps confirm that the films grow on the substrates with an initial layer of $[CoO_2]^-$, followed by the first layer of $Pd^+$ ions. While the Pd signal is well-defined, a Co signal remains detectable within the surface layers of the substrates. In the Co $L_{3,2}$ edge elemental maps in Figure 3(a) and (c), the positions of the line spectra extracted in Figure 3(b) and (d) for the Co $L_{3,2}$ edge and the corresponding O K edge are indicated by indexed arrows.

The graphs reveal that a Co signal can be detected down to the second layer of the substrates. Notably, a clear shift of the Co $L_{3,2}$ edge towards lower energies and a change in the $L_3/L_2$ edge ratio are observed upon crossing the interface to the substrate (between line signals 4 and 3). This shift and change in the edge ratio suggest a reduction in the cobalt valence state compared to the $Co^{3+}$ valence in the films. From the first $[CoO_2]^-$ layer in the films onward, the shape and energy loss of the edges remain consistent. The O K line signal 1 exhibits the typical shape of an O K edge in $Al_2O_3$ and $LaAlO_3$, respectively, while the O K edge extracted from signal 7 and signal 6 within the film displays the characteristic shape of octahedrally coordinated oxygen ions in $PdCoO_2$.

Investigations into interface reconstructions in delafossites with different A and B ions ($CuCrO_2/Al_2O_3$) have revealed the presence of a stabilizing interfacial reconstruction involving the formation of a monolayer-thick $CuCr_{1-x}Al_xO_2$ alloy. In this case, a significant reduction in the Cr $L_{3,2}$ edge signal was observed without a chemical shift or change in the edge ratio across the interface. This suggests that the interface reconstruction observed for $PdCoO_2$ is fundamentally distinct[20]. The EELS investigations indicate that, alongside chemical intermixing at the interface, the secondary phase we observe at the interface contains cobalt in a lower oxidation state compared to the film.

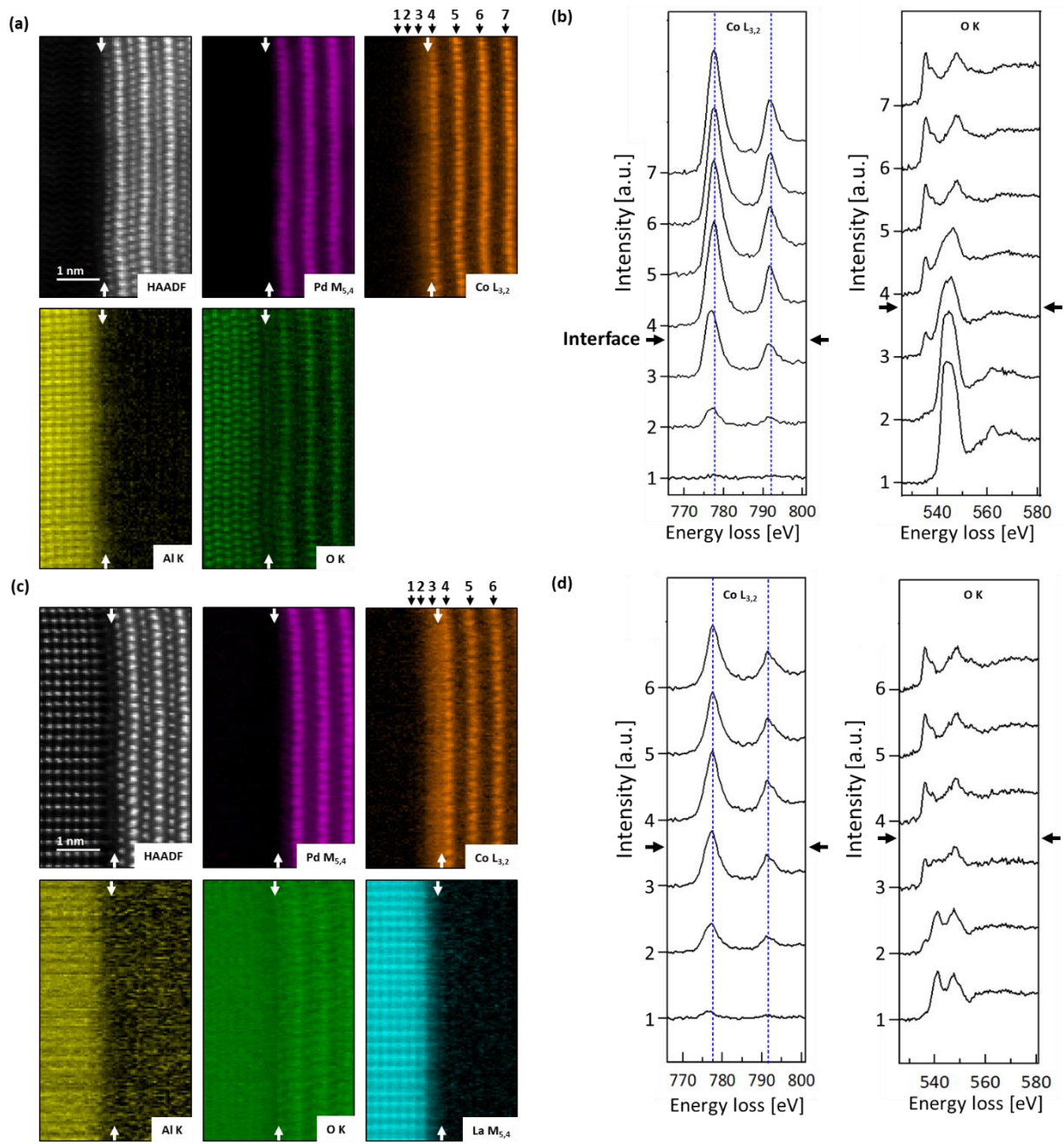

Figure 3. EELS elemental mapping of the (a) PdCoO$_2$/Al$_2$O$_3$ and (c) PdCoO$_2$/LaAlO$_3$ interfaces obtained by extracting the indicated edge signals from a 2D spectrum image along the [010] zone axis. The films commence with a layer of [CoO$_2$]$^-$, and the elemental maps reveal chemical intermixing at the interface between the substrate and the film. Notably, the sharp Pd signal indicates that the Pd layer is not significantly involved in the interfacial reconstruction. The O K edge elemental maps exhibit a minor Pd signal due to an overlap of the characteristic edges of the two elements in the EELS spectrum. White arrows indicate the physical interface between the substrate and the film. (b) and (d) display the Co L$_{3,2}$ and O K edge signals, extracted along the lines indicated by the indexed arrows in (a) and (c). A strong Co signal is detectable up to two layers deep within the substrates, accompanied by a shift of the edge towards lower energy-loss and a change in the L$_3$/L$_2$ edge ratio. The O K edge exhibits features characteristic of the bulk materials far from the interface and a mixed signal at the interface.

To precisely examine the interfacial phase on a large scale, advanced imaging methods are required to provide strong contrast for light elements (O and Al) across extensive regions. While ABF imaging provides a good initial impression of the lighter element distribution, phase-related imaging techniques in STEM offer superior contrast for both light and heavy elements simultaneously. Compared to phase contrast imaging with conventional high-resolution transmission electron microscopy (HRTEM), direct focused probe scanning techniques like single-sideband (SSB) and Wigner distribution deconvolution (WDD) exhibit a notably simpler contrast transfer function (CTF). These methods do not require aberrations to generate contrast and lack zero crossings, simplifying the imaging process significantly.

By raster-scanning the convergent electron beam across the sample, we acquired 4D STEM datasets by collecting a 256x256 pixel binary diffraction pattern for each probe position during the 2D scan. The use of the binary mode of the MerlinEM Medipix3 camera facilitates rapid acquisition at 20833 frames per second. The presence of the sample, particularly a very thin one satisfying the weak-phase object approximation (WPOA), induces a phase shift in the incoming electron wave, enabling effective imaging of the interfacial reconstruction involving Al and O[41]. Since a direct measurement of the exit-wave phase is physically impossible and only intensities can be measured, reconstruction techniques are necessary to back-calculate the phase from the diffraction patterns of the 2D scan. Each diffraction pattern in the dataset contains vital information about both the amplitude and phase of the transmitted electron wave.

A solution to this "phase problem" for in-focus 4D STEM diffraction patterns is single-sideband ptychography. The interference patterns between the central beam and the diffracted beams encode the relative phase information of transmitted and scattered electrons, crucial for reconstructing the sample's transmission function. By extracting the regions of interference (double-overlap regions) between the direct and scattered beam for each spatial frequency – regions containing phase changes induced by the sample – high signal-to-noise phase images of the structure are obtained. For a detailed description of the single-sideband phase reconstruction algorithm, we refer to previous studies[23,41].

Figure 4(a) presents the result of the single-sideband reconstruction of an in-focus 4D STEM dataset of the $PdCoO_2/Al_2O_3$ interface. The bottom shows the $Al_2O_3$ substrate, and the top shows the delafossite thin film, initiated with a $[CoO_2]^-$ octahedral layer, followed by the $Pd^+$ layer along the [010] zone axis, grown at a 30° rotation relative to the substrate orientation. In the second $[CoO_2]^-$ layer, the orientation of the $CoO_6$ octahedra is mirrored to the first layer in the film. A twin domain is present here, likely arising from the energetically equivalent growth of the two orientations. At the interface, the atomic reconstruction on the substrate surface is clearly observed, with repeating features occurring at regular intervals, indicated by white ticks. Figure 4(b) shows a higher magnification of one such feature, revealing three distinct regions within each feature as indicated by the insets. Given that the sample, albeit very thin, is still a three-dimensional structure, these regions can be attributed either to areas of CoO from 180° opposing orientations or to areas where both models must be superimposed with a 180° swap of the octahedral orientation.

Figure 4(c) shows the single-sideband reconstruction from an in-focus 4D STEM dataset of the $PdCoO_2$/$LaAlO_3$ interface for the [010] orientation of the film. At the interface, the atomic reconstruction in the substrate surface is observed, with repeating features occurring at regular intervals, indicated by white ticks. These features are similar to the ones of the $PdCoO_2$/$Al_2O_3$ interface, but with a higher periodicity, as $PdCoO_2$ has a stronger lattice mismatch with $LaAlO_3$ (-9.06 %) compared to $Al_2O_3$ (+2.9 %). Details on the lattice mismatch calculations for the film/substrate combinations investigated in tis work can be found in the Supplementary Material. The insets in Figure 4(c) also suggest the presence of regions of single CoO orientation and regions where both CoO models are superimposed with a 180° swap of the octahedral orientation. Interestingly, we observe a similar reconstruction at the interface for both systems, despite the difference in lattice parameter relationships: $a_{film} > a_{substrate}$ in $PdCoO_2$/$Al_2O_3$ and $a_{film} < a_{substrate}$ in $PdCoO_2$/$LaAlO_3$. While the films are relaxed in the final state, these lattice relationships suggest that, in the absence of relaxation mechanisms, compressive and tensile strain would initially be expected upon deposition respectively.

To identify the driving force behind the octahedral distortion, we performed phase reconstructions along the [210] orientation of the films, where the octahedra lie in projection, i.e., opposing orientations cannot be distinguished. Supplementary Figure S4 shows 4D STEM phase reconstructions of the (a) $PdCoO_2$/$Al_2O_3$ interface and (b) $PdCoO_2$/$LaAlO_3$ interface. For both images, we analyzed the lattice modulation using the phase lock-in method developed by Goodge et al.[42]. Through this method, we extracted the amplitude and phase of a lattice peak associated with the film lattice, indicated by the yellow spot in the Fourier transform insets. At the interface, within the interfacial phase, we observe defects that manifest in a $2\pi$ wrap-around of the modulation phase. Notably, due to the higher periodicity, several such defects can be identified in the field of view for the $LaAlO_3$ interfacial reconstruction. These defects are likely responsible for the octahedral distortion at the interface and compensate the mismatch strain between film and substrate. Combined with the previous findings, this provides a detailed picture of the interfacial reconstruction mechanism that stabilizes $PdCoO_2$ thin films on $Al_2O_3$ and $LaAlO_3$ substrates.

The repeating features indicate that a thin CoO secondary phase forms within the substrate surface during growth. EELS elemental mapping for both substrates, along with fine-structure analyses showing a change in the Co valence across the interface towards lower values, confirm the presence of Co in the upper layers of the substrate. Since cobalt exhibits a valence of $2^+$ in bulk CoO in contrast to $3^+$ in $PdCoO_2$, this strongly supports the presence of distorted CoO at the interface. This conclusion is further supported by the fit between the atomic models and phase reconstructions.

The presence of CoO within the substrate surface leads to a novel understanding of the mechanism that relieves the lattice mismatch for the film, contrary to the traditional understanding of the relief mechanisms at oxide heterostructure interfaces. In general, it is observed that the lattice misfit at a coherent interface is low and accommodated by the elastic deformation of neighboring lattices, resulting in a nearly perfect match between atoms at the interface. At a semi-coherent interface it is moderate and compensated by the formation of a periodic array of interfacial misfit dislocations. Finally, the lattice misfit at incoherent interfaces is very large and adjacent crystals on both sides of the interface retain their original lattice and are rigidly stacked against each other, making it difficult for interfacial misfit dislocations to form[43]. The formation of a separate phase of the film material within the substrate is not part of these standard mechanisms, however, similar formation of secondary phases has been observed in other heterojunction systems. For instance, van der Waals interfacial reconstructions between transition metal dichalcogenides (TMDs) and gold have been reported, involving the formation of a metastable $AuS_4$ phase at the interface[44]. Similarly, in hexagonal $LuFeO_3$ intergrown with $Fe_3O_4$ nanolayers, a distinct interfacial rearrangement occurs, stabilizing the nanolayered ferrite in a process recognized as a universal mechanism for polar surfaces[45]. In more complex oxide heterostructures, interfacial reconstructions drive the formation of entirely new phases. In the case of a two-dimensional electron gas (2DEG) at the $TiO_2/LaAlO_3$ interface, oxygen vacancies in anatase $TiO_2$ promote the emergence of an additional alloyed $TiAlO_4$ layer. STEM studies have provided direct evidence of this extra layer at the earliest stages of $TiO_2$ growth [37]. Likewise, CaO films grown on Mo(001) exhibit Mo diffusion from the substrate into the film, replacing approximately 25% of the Ca ions leading to the formation of a rocksalt-type $CaMoO_4$ structure. This substitution mechanism facilitates the development of extended, defect-free oxide patches[46]. More generally, interface reaction mechanisms involving topotactic reactions have been studied as a less common means for epitaxial systems to accommodate lattice misfit at reactive spinel interfaces[47] These findings underscore that interfacial reconstructions and emergent secondary phases are not merely byproducts of growth but intrinsic processes that can drive the stabilization and transformation of materials, often leading to novel structural and electronic properties.

The requirement of specifically a CoO phase at the interface between $PdCoO_2$ and $Al_2O_3/LaAlO_3$ is further supported by the fact that we observed the films to completely decompose without nucleation when the growth is initiated with a Pd layer, as indicated by DFT calculations, which show an energetically more stable $Al_2O_3/[CoO_2]^-$ interface than an $Al_2O_3/Pd^+$ interface (see Supplementary Figure S2).

To further investigate interfacial stability, we performed DFT calculations incorporating an additional CoO layer at the interface, effectively resulting in a Co bilayer of CoO(111). However, in the absence of interfacial reconstructions, this Co-bilayer experiences a 10% in-plane compressive strain relative to bulk CoO(111) on the $Al_2O_3$ substrate. As a result, geometry optimizations for the additional CoO layer also led to unphysically large out-of-plane bond elongations, similar to the previous case of Co incorporation into $Al_2O_3$. This strongly indicates that a secondary CoO phase at the interface cannot exist in its ideal bulk-like form without undergoing structural reconstruction, a conclusion corroborated by STEM investigations. While DFT is a powerful tool for modeling interfacial reconstructions, directly simulating the experimentally observed reconstructions without prior knowledge of a specific structural candidate is computationally unfeasible due to the excessive size of the required supercells.

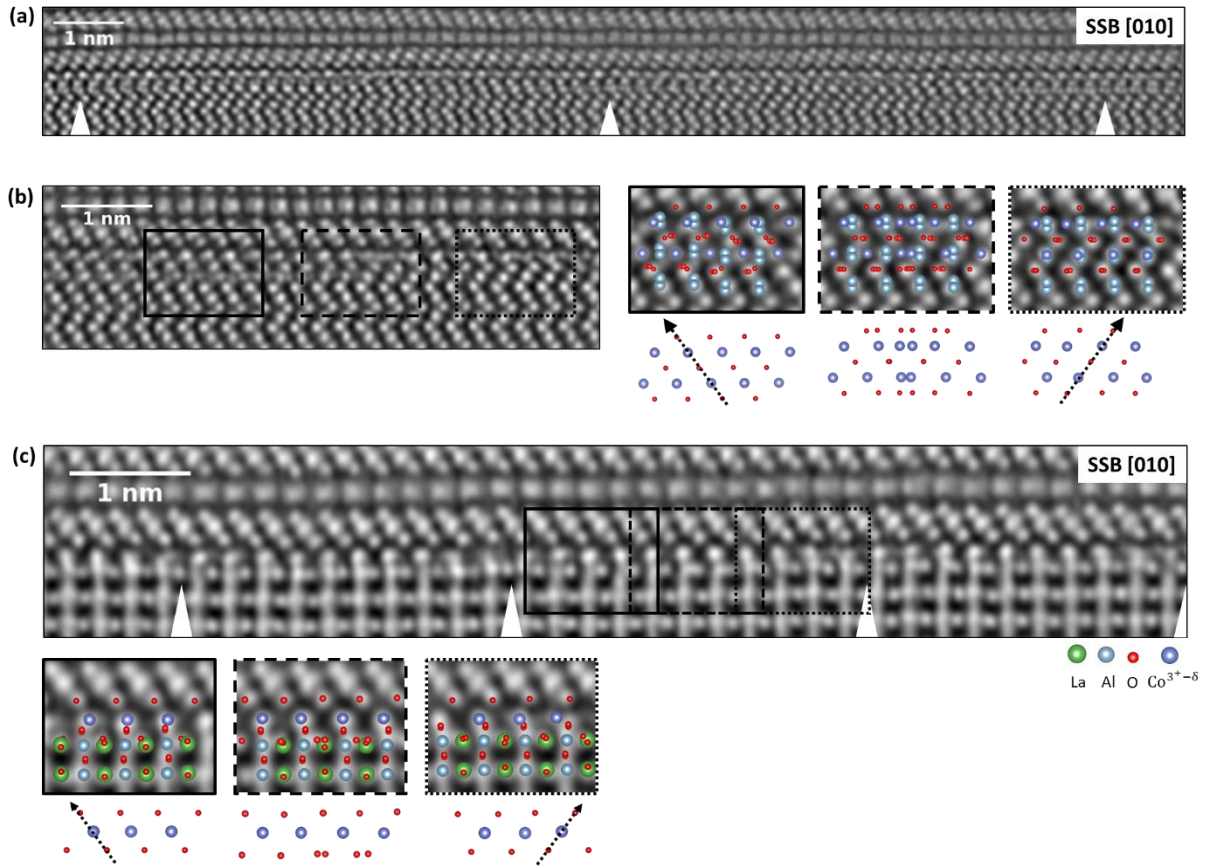

Figure 4. (a) Single-sideband ptychographic phase reconstruction from a 4D STEM dataset with 90x800 probe positions acquired across the PdCoO$_2$/Al$_2$O$_3$ interface. The reconstruction provides strong contrast for all elements and reveals the presence of periodically repeating features within the substrate surface layer, indicated by white ticks. (b) High-magnification single-sideband reconstruction of the interface revealing the presence of subdomains, indicated by black boxes, with two opposing octahedral orientations of the CoO phase. (c) Single-sideband ptychographic phase reconstruction from a 4D STEM dataset with 100x600 probe positions acquired across the PdCoO$_2$/LaAlO$_3$ interface. The insets reveal subdomains with two opposing octahedral orientations of the CoO phase.

In addition to Al$_2$O$_3$ and LaAlO$_3$, we attempted to grow PdCoO$_2$ thin films on (111) SrTiO$_3$ substrates via MBE. Despite a lattice mismatch of -11.04 %, similar to that of PdCoO$_2$ on LaAlO$_3$ (-9.06%), analysis of the film structure in Supplementary Figure S5 revealed that the film is almost entirely decomposed, indicating that stable growth is essentially impossible. In the few areas where a film does grow (blue inset in Figure S5), it can be observed that instead of growing with an energetically favorable 30° rotation on the substrate (green frame in Supplementary Figure S5), the film grows with a 60° rotation (blue frame in Figure S5). We conclude that the complex interplay between Al in the Al$_2$O$_3$ and LaAlO$_3$ substrates and Co at the interface facilitates the formation of the interfacial phase, contributing to stable film growth of PdCoO$_2$. This behavior is not observed for PdCoO$_2$ on SrTiO$_3$.

To exclude the influence of the growth mode on the formation of the interfacial phase, we fabricated PdCoO$_2$ films on Al$_2$O$_3$ using pulsed laser deposition (PLD). As shown by the SSB phase reconstructions for two zone axes and the structural models at the interface in Supplementary Figure S6, we again observed the presence of the interfacial phase in the substrate surface. This strongly suggests that this phase forms regardless of the chosen fabrication method.

Having fully understood the growth mechanism and the stabilization of the PdCoO$_2$ epitaxial film on Al$_2$O$_3$ and LaAlO$_3$ substrates, we are now able to grow films that would otherwise not exhibit stable growth on Al$_2$O$_3$. This is achieved by first nucleating a few-unit-cell thick buffer layer of high-quality PdCoO$_2$ under optimized growth conditions[48]. Further studies on delafossites with other *A* and *B* cations have demonstrated the effectiveness of a stable delafossite buffer layer for the epitaxial growth of other delafossites. For instance, CuCrO$_2$ buffer layers have been successfully employed for the growth of PdCrO$_2$[17,20].

# Conclusions

We present insights into a unique atomic reconstruction at the interface between oxide MBE-grown PdCoO$_2$ thin films and Al-containing substrates Al$_2$O$_3$ and LaAlO$_3$. This reconstruction deviates from the traditional understanding of relief mechanisms at oxide heterostructure interfaces. It is characterized by unmatched atomic columns within the substrate's surface layers, suggesting the formation of an ultra-thin secondary phase. Extensive STEM analyses, including ptychographic phase reconstructions and STEM-EELS elemental mapping, indicate the presence of a CoO phase at the interface modulating periodically within the substrate, with cobalt existing in a lower valence state (Co$^{3+-\delta}$) compared to the Co$^{3+}$ state in PdCoO$_2$. Observations of unstable growth of PdCoO$_2$ when initiated with a Pd$^+$ layer emphasize the crucial role of this interfacial phase in enabling successful epitaxial growth.

These findings underscore the complex interplay between the elements of film and substrate during growth, where atomic-scale defects and valence changes at the interface significantly influence the structural properties of the thin films. Understanding these phenomena is essential for optimizing growth processes. Furthermore, utilizing high-quality PdCoO$_2$ buffer layers offers a novel pathway for the stable synthesis of previously unachievable delafossite thin films.

# Acknowledgments

The authors would like to thank Marion Kelsch and Ute Salzberger for their excellent work in the preparation of the TEM samples.

The synthesis work at Cornell was primarily supported by the U.S. Department of Energy, Office of Basic Sciences, Division of Materials Science and Engineering under Award No. DE-SC0002334. This research was funded in part by the Gordon and Betty Moore Foundation's EPiQS Initiative (Grant No. GBMF9073 to Cornell University). Substrate preparation was performed, in part, at the Cornell NanoScale Facility, a member of the National Nanotechnology Coordinated Infrastructure, which is supported by the NSF (Grant No. NNCI-2025233).

We thank Roland Gillen for fruitful discussions. The authors gratefully acknowledge the scientific support and HPC resources provided by the Erlangen National High Performance Computing Center (NHR@FAU) of the Friedrich-Alexander-Universität Erlangen-Nürnberg (FAU).

# Supplementary Material:

# Unveiling the Interfacial Reconstruction Mechanism Enabling Stable Growth of the Delafossite PdCoO$_2$ on Al$_2$O$_3$ and LaAlO$_3$

Anna Scheid[1], Tobias Heil[1], Y. Eren Suyolcu[1], Qi Song[2], Niklas Enderlein[3], Arnaud P. Nono Tchiomo[4], Prosper Ngabonziga[4,5], Philipp Hansmann[3], Darrell G. Schlom[2,6,7], and Peter A. van Aken[1]

[1]Max Planck Institute for Solid State Research, Stuttgart, 70569, Germany

[2]Department of Materials Sciences and Engineering, Cornell University, Ithaca, New York 14853, USA

[3]Department of Physics, Friedrich-Alexander-Universität Erlangen-Nürnberg (FAU), 91058, Erlangen, Germany

[4]Department of Physics and Astronomy, Louisiana State University, Baton Rouge, Louisiana 70803, USA

[5]Department of Physics, University of Johannesburg, P.O. Box 524 Auckland Park 2006, Johannesburg, South Africa

[6]Kavli Institute at Cornell for Nanoscale Science, Ithaca, New York 14853, USA

[7]Leibniz-Institut für Kristallzüchtung, Berlin, 12489, Germany

Thin films of PdCoO$_2$ were synthesized by shutter-controlled MBE in a Veeco Gen10 MBE system on (001) sapphire and (111)$_{pc}$ LaAlO$_3$ substrates. Details on the film growth can be found in the Materials and Methods section and the Supplementary Material of a previous study[1].

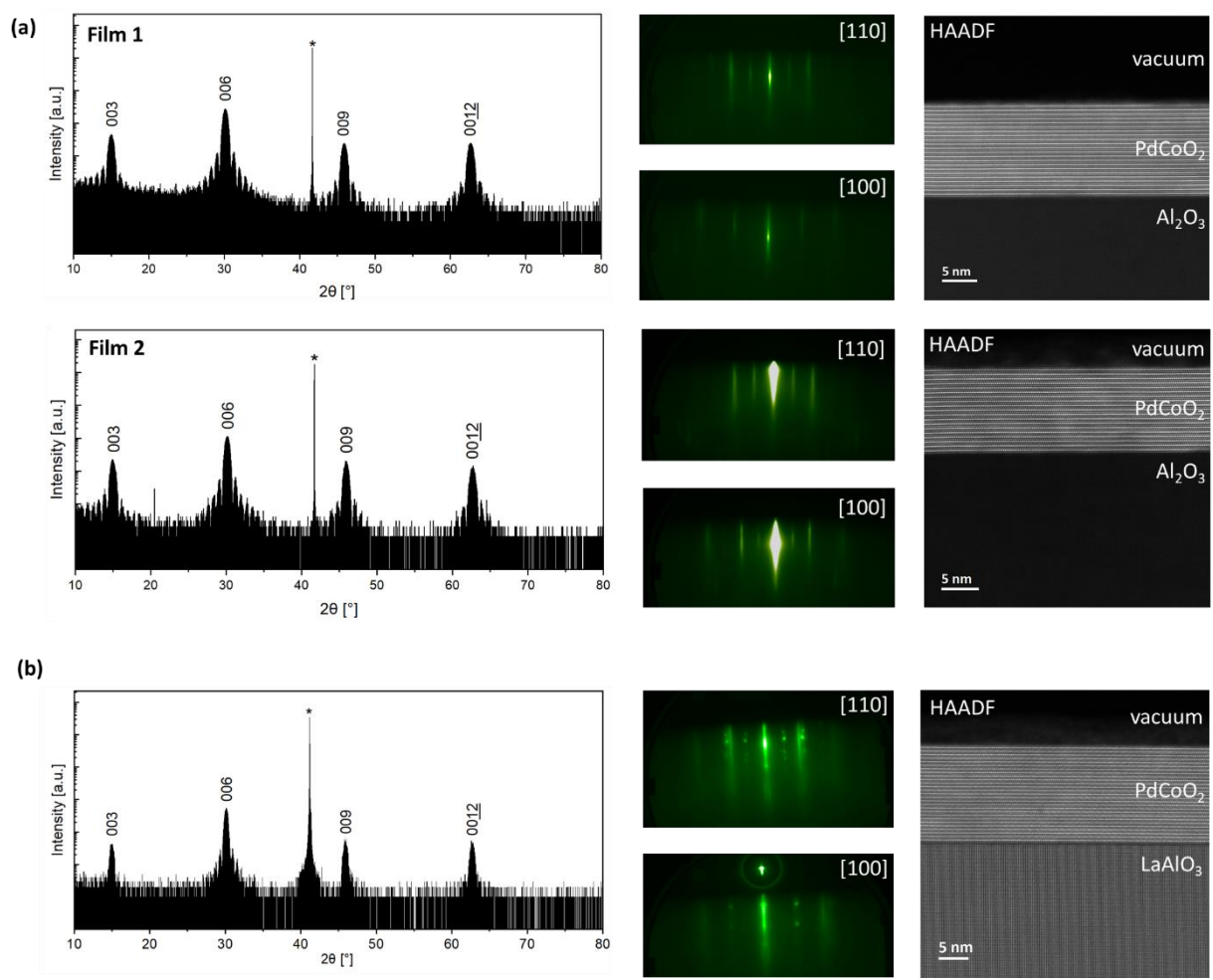

Figure S1. (a) X-ray diffraction, in-situ RHEED patterns, and STEM HAADF images of two PdCoO$_2$ thin films grown on an Al$_2$O$_3$ substrate. The Laue oscillations observed for the PdCoO$_2$ peaks indicate a smooth film-substrate interface, which is also evident in the low-magnification STEM images. (b) X-ray diffraction, in-situ RHEED patterns, and STEM HAADF images of a PdCoO$_2$ thin film grown on LaAlO$_3$. Asterisks (∗) denote substrate reflections.

# Density Functional Theory (DFT) calculations

DFT calculations were performed using the Quantum ESPRESSO suite[2–4] to assess the likelihood of different PdCoO$_2$/Al$_2$O$_3$ interface types. The interface can be most effectively understood by a shared structural feature in the oxygen network: both structures exhibit coinciding oxygen triangles at the interface (yellow triangles in Figure S2(a)), serving as the covalently bonding bridge between the two materials. To construct the DFT unit cells for the interface simulation, rectangular in-plane supercells of both materials were generated, containing an equal number of oxygen triangles. The unit cell vectors $(\vec{a}_S, \vec{b}_S)$ correspond to linear combinations of the primitive cell vectors $(\vec{a}_P, \vec{b}_P)$. For PdCoO$_2$, the chosen supercell was spanned by $\vec{a}_S = 2\vec{a}_P + \vec{b}_P$ & $\vec{b}_S = 3\vec{b}_P$ and for Al$_2$O$_3$ by $\vec{a}_S = \vec{a}_P$ & $\vec{b}_S = \vec{a}_P + 2\vec{b}_P$, respectively.

As highlighted in Figure S2(a), the rectangular cells feature two/three oxygen triangles along the short/long side of the rectangle, respectively. To accurately reflect the experimental setup, the in-plane lattice constants of the rectangular simulation cells were fixed to the experimentally determined lattice constant of Al$_2$O$_3$. Consequently, the lattice spacing of PdCoO$_2$ (being 2.98 % larger) is slightly compressed compared to its bulk value. For DFT calculations, the construction of periodically continued unit cells is necessary, effectively resulting in a stacked, alternating structure with two identical interfaces per unit cell (see respective unit cell at the bottom of Figure S2(b)). To optimize the structure, we relaxed the lattice constant along the interface normal vector and the atomic positions of all atoms within the unit cell.

The stability of different interfaces was estimated by comparing the total energy of the (optimized) interface and that of its two constituents (see Figure S2(b) - note that the factor two (2Δ) in the equation accounts for the fact that there are two identical interfaces in the simulation cell)[5,6]. The latter was computed in two separate self-consistent field (SCF) calculations from the deconstructed unit cell without any additional geometry optimization. Since the oxygen layer at the interface is shared by both components (Al$_2$O$_3$ and PdCoO$_2$), there are, in principle, two ways to define the isolated constituents from the full simulation cell for the calculation of the interface energy (according to the scheme in Figure S2 (b)): the shared oxygen layer can be ascribed to either Al$_2$O$_3$ or PdCoO$_2$.

The interface energies were calculated using both ascriptions. The energetic differences between the various interfaces are represented as matrices in Figure S2(c). When the O-layer is ascribed to Al$_2$O$_3$ (i.e., the interface energy is defined as the binding energy between the shared O-layer and the Co- or Pd-layer, respectively), we find that the Co-interface is significantly more stable (by 4.0 eV per unit cell area of ~40 Å$^2$). This is reasonable considering that Pd forms covalent bonds with only two oxygen atoms at a 180° bond angle, while Co bonds with six oxygen atoms to form an octahedron (see Figure S2(b)). Consequently, the Co-interface contains three times more Co-O bonds than the Pd-interface has Pd-O bonds, resulting in greater binding energy.

Conversely, in the Pd-interface, each oxygen atom in the shared O-layer is bonded to only one Pd atom, whereas in the Co-interface, each oxygen atom is bonded to three Co atoms. Thus, the shared O-layer in the Pd-interface exhibits a higher demand for electrons from the Al-layer, as the Pd-interface donates less electron density to the shared O-layer compared to the Co-interface. This suggests that the shared O-layer is more strongly bonded to the first Al-layer of $Al_2O_3$ in the Pd-interface than in the Co-interface, which is confirmed by the binding energies obtained from ascribing the shared O-layer to the delafossite. This can be seen as a direct consequence of the Co-interface being more stable than the Pd-interface in terms of their bonding strength to the shared O-layer.

In addition, different cases of Co-substitution in the first Al-layer of the substrate were tested. The results, presented as 50% Co-substitution (affecting two of the four Al-atoms per Al-layer) and the 100% case, are shown in the antisymmetric matrix in Figure S2(c).

## Technical details

For the DFT calculations performed with the Quantum ESPRESSO suite[2–4], optimized norm-conserving pseudopotentials[7,8] were used with an energy cutoff of 120 Ry, and the exchange-correlation interaction was approximated by the Perdew-Burke-Ernzerhof functional (PBE). Brillouin zone integration was carried out on a 6×3×1 electron momentum grid, which corresponds to an equal sampling density for the rectangular supercell with the in-plane lattice constants $a$=4.7606 Å and $b$=8.2456 Å. The out-of-plane lattice constant $c$ was larger than 30 Å for all considered simulation cells, such that using only one $k$-point along the $k_z$ direction is sufficient. A Gaussian smearing of 0.02 Ry was included. The derived total energies were converged within the accuracy of 0.03 eV for the chosen numerical parameters.

During the geometry optimizations, the mentioned in-plane lattice constants, which correspond to the experimental lattice spacing of $Al_2O_3$, were kept fixed, while the out-of-plane lattice constant (i.e., the stacking direction) and the atomic positions were fully optimized. Different thicknesses of the $Al_2O_3$ compound in the simulation cells were tested, namely 4, 6, and 8 Al layers. After optimizing these respective simulation cells with different thicknesses of the $Al_2O_3$ part, the variation in the calculated interface energy was found to be less than 0.01 eV. This confirms that the chosen layer thicknesses (of at least 6 Al layers in the pure, i.e., unsubstituted $Al_2O_3$) are sufficient to achieve robust and reliable results.

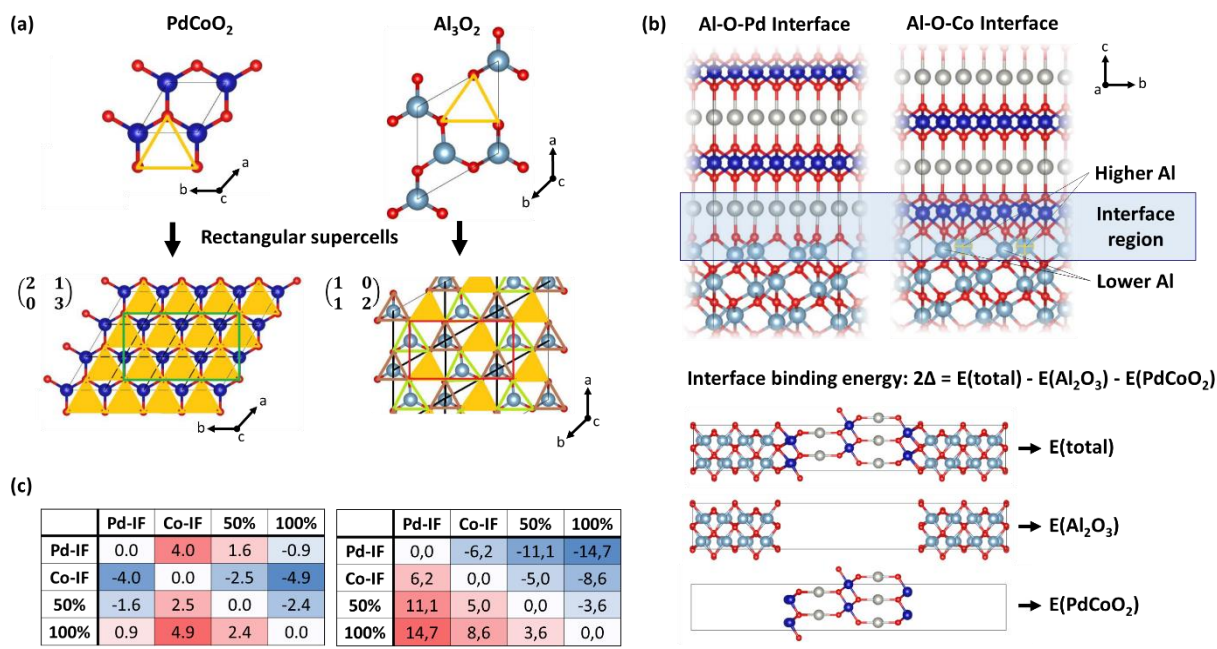

Figure S2. (a) Oxygen layers exhibiting a triangular (highlighted) in-plane arrangement, a shared structural feature between $PdCoO_2$ and $Al_2O_3$. The top panel displays the primitive hexagonal unit cells of the two interfaced compounds, while the bottom panel shows the specific rectangular supercells utilized for the interface simulations. (b) Top panel: Side views of the Pd- and Co-interface simulation geometries. The interface region separating the two compounds is highlighted with transparent blue. Co-substituted Al-sites are highlighted as 'Higher Al' and 'Lower Al' depending on their $c$-axis position. The interface binding energy is calculated as the difference between the total energy of the (optimized) interface and that of its two constituents. (c) Differences in calculated interface energies (per unit cell area of ~40 Å$^2$) between the (in order) Pd-, Co-, 50% Co-substituted, and 100% Co-substituted interfaces. For 50% Co-substitution, various substitution patterns were tested and averaged in the table for simplicity, as all were energetically positioned between the extreme cases of 0% and 100%. The differences are represented as an antisymmetric matrix, where a negative matrix element (i, j) indicates that the i-interface (row index) is more stable than the j-interface (column index). Matrix elements are color-coded: negative values are shown in blue, positive values in red, with a smooth interpolation between the extreme values. Zero values (on the diagonal) are represented in white. The table on the left corresponds to the calculation of interface energies with the shared oxygen layer attributed to $Al_2O_3$, and the table on the right to the delafossite ($PdCoO_2$) constituent of the simulation cell, respectively.

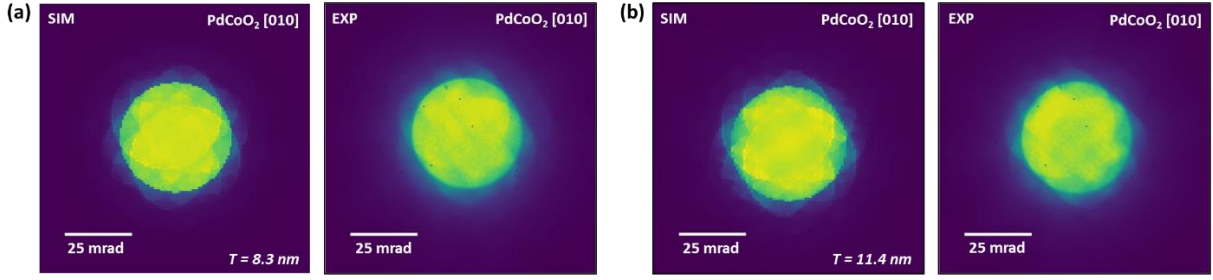

Figure S3. Simulated (SIM) and experimental (EXP) PACBEDs for the [010] zone axis of $PdCoO_2$ from 4D STEM scans across the (a) $PdCoO_2/Al_2O_3$ and (b) $PdCoO_2/LaAlO_3$ interface.

## Lattice mismatch calculations

The lattice mismatch quantifies the relationship between the crystalline lattice parameters at the film–substrate interface. In epitaxial thin films, it relates the lattice parameters of the relaxed film and substrate. According to Frank and van der Merwe, the lattice mismatch is given by equation (1)[9]. For delafossites, growing with a 30° rotation on the substrate, the in-plane lattice parameter of the film must be corrected by a factor of cos(30°) at the interface.

$$\varepsilon_\mathrm{m} = \frac{a_{film} - a_{substrate}}{a_{substrate}} \quad (1)$$

As a result, for negative values of $\varepsilon_\mathrm{m}$, i.e. $a_{film} < a_{substrate}$, the films are under tensile strain when growth is initiated and for positive values of $\varepsilon_\mathrm{m}$, i.e. $a_{film} > a_{substrate}$, the films nucleate under compressive strain on the substrate.

Based on the sbstrate and film lattice parameters listed below, the following lattice mismatches can be calculated for the systems investigated in this work:

|  | $PdCoO_2/Al_2O_3(001)$ | $PdCoO_2/LaAlO_3(111)_{pc}$ | $PdCoO_2/SrTiO_3(111)_{pc}$ |
|---|---|---|---|
| $a_{film}$ [Å] | cos(30°)·5.66 | cos(30°)·5.66 | cos(30°)·5.66 |
| $a_{substrate}$ [Å] | 4.76 | 5.39 | 5.51 |
| $\varepsilon_\mathrm{m}$ | +2.98 % | -9.06 % | -11.04 % |

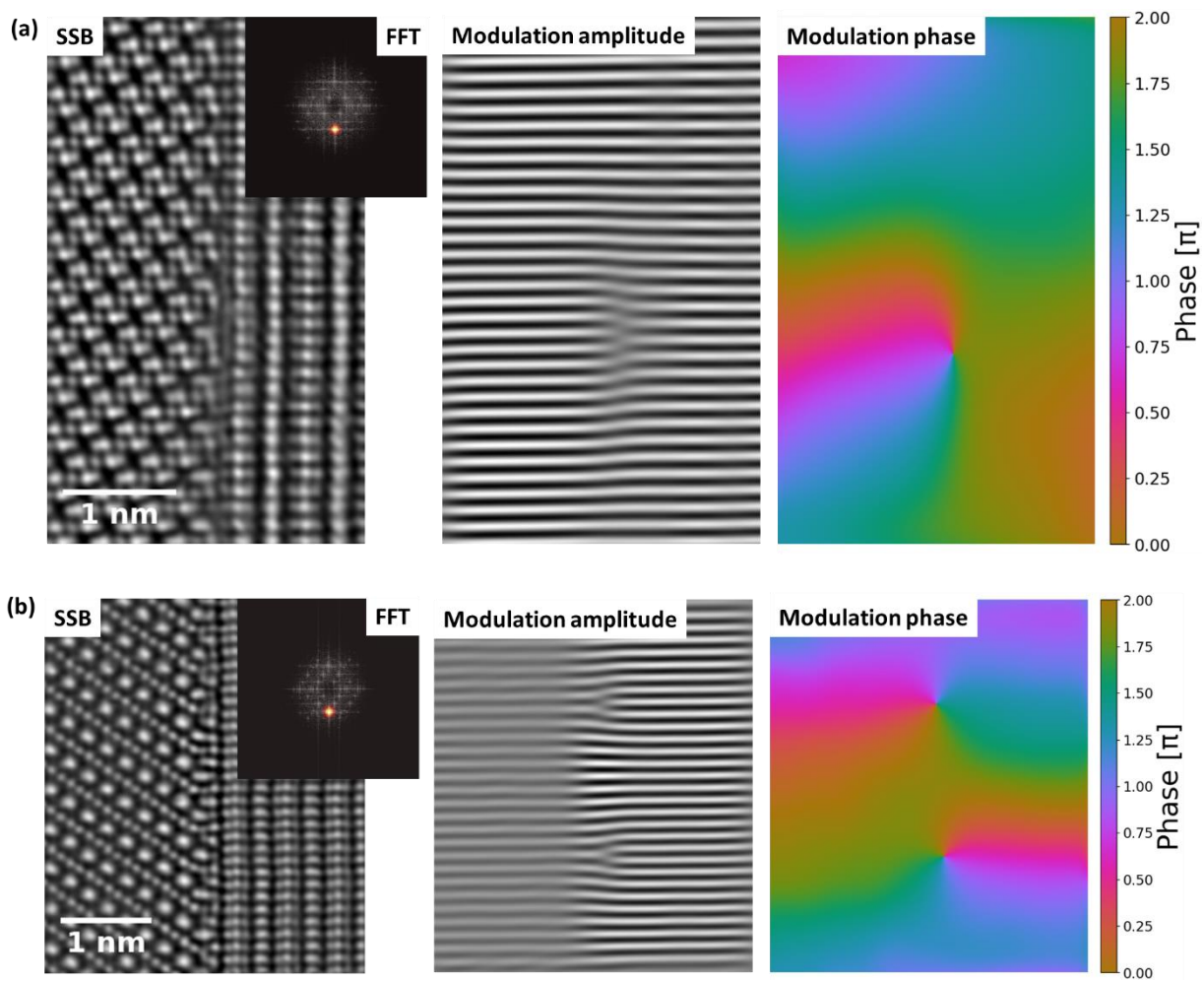

Figure S4. Phase lock-in lattice-modulation analysis from single-sideband phase reconstructions along the [210] zone axis of $PdCoO_2$ on (a) $Al_2O_3$ and (b) $LaAlO_3$. Extracted modulation amplitude and phase from the fast Fourier transformed image reveal the presence of defects in the interfacial phase.

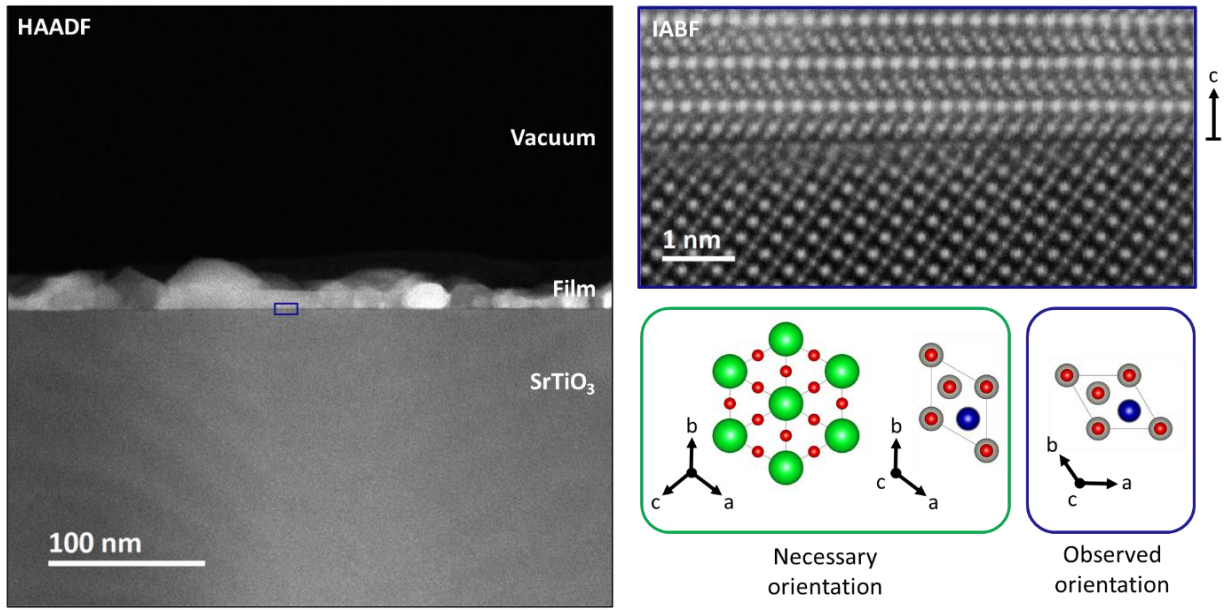

Figure S5. HAADF STEM image of a decomposed $PdCoO_2$ thin film grown on a (111) $SrTiO_3$ substrate. The blue inset highlights an area where decomposition is minimal, revealing a $PdCoO_2$ thin film grown with a 60° rotation instead of the expected 30° rotation on the substrate, as observed in the inverted annular bright-field (IABF) image. The green frame indicates the expected film orientation relative to the (111) $SrTiO_3$ substrate for stable growth, while the blue frame shows the observed orientation along the c-axis of the film.

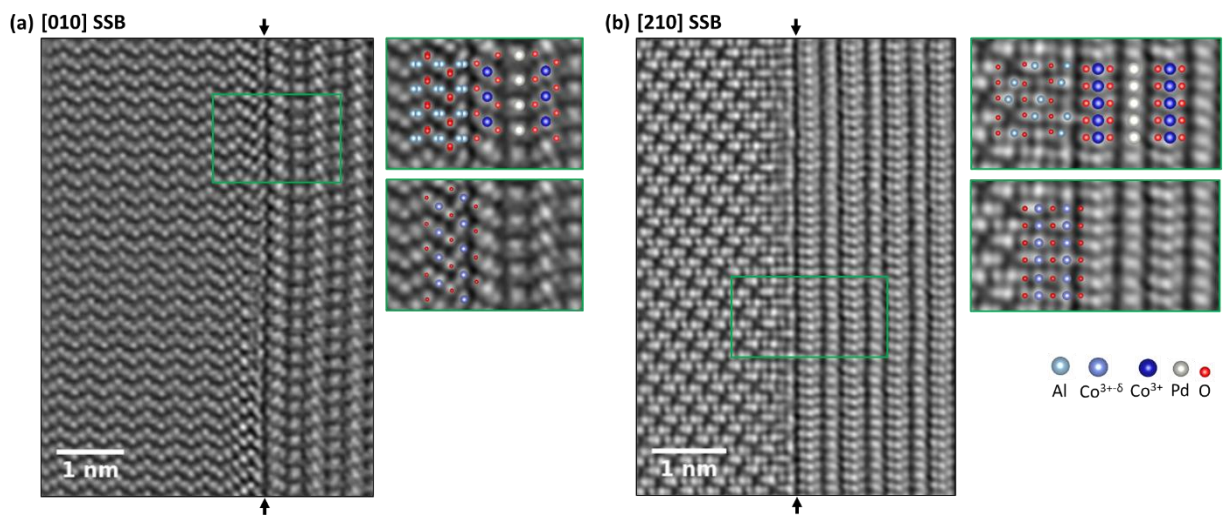

Figure S6. Single-sideband ptychographic phase reconstructions from 4D STEM datasets acquired across the $PdCoO_2/Al_2O_3$ interface for the PLD-grown thin films along the (a) [010] and (b) [210] film orientations. The reconstructions provide strong contrast for all elements and reveal the presence of periodically repeating features within the substrate surface layer. Insets show structural model overlays with the film, substrate, and interfacial phase models along the respective zone axes.